\documentclass[pre,aps,showpacs,twocolumn]{revtex4-1}

\usepackage{graphicx}
\usepackage{pstricks}
\usepackage{amssymb}
\usepackage{amsmath}
\usepackage{amsfonts}
\usepackage{natbib}

\graphicspath{{figure/}}
\newcommand{\beq}{\begin{equation}}
\newcommand{\eeq}{\end{equation}}
\newcommand{\bea}{\begin{eqnarray}}
\newcommand{\eea}{\end{eqnarray}}

\begin{document}
%
%

\title {Energy versus information based estimations of dissipation 
using a pair of magnetic colloidal particles}
\author{S. Tusch$^2$, A. Kundu$^1$, G. Verley$^1$, T. Blondel$^1$, 
V. Miralles$^2$, D. D{\'e}moulin$^2$, D. Lacoste$^1$, J. Baudry$^2$}
\affiliation{$^1$ Laboratoire de Physico-Chimie Th\'eorique - UMR CNRS Gulliver 7083,
ESPCI, 10 rue Vauquelin, F-75231 Paris, France}
\affiliation{$^2$ Laboratoire LCMD, ESPCI, 10 rue Vauquelin, F-75231 Paris, France}
\date{\today}

\begin{abstract}
Using the framework of stochastic thermodynamics, we present an experimental study of a doublet 
of magnetic colloidal particles which is manipulated by a time-dependent magnetic field. 
Due to hydrodynamic interactions, each bead experiences 
a state-dependent friction, which we characterize using a hydrodynamic model. 
In this work, we compare two estimates of the dissipation in this system: the first one is energy 
 based since it relies on the measured interaction potential, while the second one is information based since it  
uses only the information content of the trajectories. While the latter only offers a lower bound of the former, 
we find it to be simple to implement and of general applicability to more complex systems.
\end{abstract}

\maketitle

In the last decade, a broad number of works have significantly 
improved our understanding of the thermodynamics of small systems. A central idea, namely 
the application of thermodynamics at the level of trajectories, has
developed into a field of its own now called stochastic thermodynamics 
\cite{Seifert2012,Jarzynski2011_vol2,Ciliberto2010_vol,Ritort2008_vol137}.
Manipulated colloids are a paradigmatic example of stochastic thermodynamics because of the ease with which colloids 
can be manipulated and observed. 

Many studies of such systems have used a single colloidal particle, in 
an harmonic \cite{Imparato2007_vol76} or anharmonic potential 
\cite{Ciliberto2010_vol,Blickle2006_vol96}, which is described by an overdamped Langevin equation with 
a constant diffusion coefficient. 
Recently, Celani et al. have pointed out that the overdamped Langevin description 
fails to capture some aspects of the thermodynamics of this system in the presence of    
multiplicative noise due to temperature gradients \cite{Celani2012}. 
In soft matter systems, temperature gradients are difficult to control at 
the micron scale but multiplicative noise arises frequently   
due to hydrodynamic friction.  
In this paper, we study such a case using a pair of magnetic colloids which
are manipulated by a time-dependent magnetic field. This system
offers a convenient mean to measure forces in various soft matter and biological systems 
because the colloids can be embedded in complex fluids or molecules of interest can be grafted on them \cite{Brangbour2011}.

In this paper, we focus on a pair of bare manipulated colloids in water. 
In the first part, we evaluate the work distribution in this system  
within stochastic thermodynamics. In the second part, we evaluate an
information theoretic bound for the dissipation in this process 
using only trajectory information.

The projection of the Brownian motion of both beads is observed 
in the plane parallel to the bottom wall with video-microscopy. 
We assume that the fluctuations perpendicular to the wall are negligible 
since the beads have settled under gravity. 
Therefore, we focus on the 
2D relative displacement vector in polar coordinates ${\bf r}=(r,\theta)$ as shown in figure 1.

The interaction between the beads is modeled using a potential, which is the sum of three contributions: the 
dipolar interaction of the magnetic beads with each 
other $U_{dip}$, the interaction $U_{mag}$ of the beads with the applied magnetic field ${\bf B}=B \hat z$, 
and a repulsive interaction of electrostatic origin $U_{el}$: 
\beq
U(r,\theta,B)=U_{dip}(B,r,\theta) + U_{mag}(B) + U_{el}(r).
\eeq
This potential has a short range repulsive part due to electrostatics and
a long-range attractive part due to dipolar interactions as described
 in \cite{Lacoste2009_vol80} and in Suppl. Mat. 
First, we prepare the system in an equilibrium state
in a constant magnetic field.
In this case, the distribution of the relative coordinate should follow a 
Boltzmann distribution, which we use to test our model.
We manage to obtain a very good fit of the data in a rather large range of magnetic field 
from $B_1=0.15$mT to $B_2=0.45$mT as shown in figure 2.
In this range, we can assume that the magnetic dipole 
moments carried by the beads have a fixed orientation along $\hat z$.
We can observe in figure 2 that 
the potential is anharmonic at low magnetic field but harmonic at high field, where the motion
of the beads becomes more confined to the vicinity of the minimum of the potential.  

Having well characterized the fluctuations of this system at equilibrium, we now  
investigate the non-equilibrium fluctuations of the same beads 
when they are driven by a time-dependent magnetic field.
The protocol of magnetic field is a periodic function of period $\tau+\tau_{eq}$, 
with $\tau$ and $\tau_{eq}$ defined in figure 1.
The explicit time dependence of the protocol is
$B(t)=B_2+(B_1-B_2) \sin^2( \pi t/\tau) $, for $0 \leq t \leq \tau$, and
$B(t)=B_2$ for $\tau \leq t \leq \tau+\tau_{eq}$. 
The time $\tau_{eq}$ represents the duration of a pause which is needed to prepare
the system at equilibrium for the beginning of the next cycle.
\begin{figure}[t!]
\includegraphics[scale=0.45]{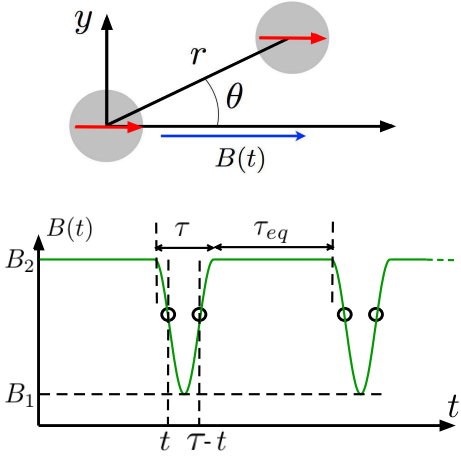}
\hspace{2mm}
\caption{\label{fig1}
Sketch of the experiment showing the two magnetic beads (top) and 
time-dependence of the applied magnetic field $B(t)$ (bottom). The protocol is
composed of driving phases of duration $\tau$ followed by pauses of duration $\tau_{eq}$.
The open circles represent two symmetric points at times $t$ and $\tau-t$ in this protocol.}
\end{figure}

We focus on the dynamics of the displacement vector ${\bf r}=(r,\theta)$, 
which we describe with the following overdamped Langevin equations:
\bea
\label{Langevin}
\Gamma_r(r) \dot{r} &=& f_r - \partial_r U(r, \theta,B) + \eta_r \circ \sqrt{2 k_B T \Gamma_r(r) }, \\
\Gamma_\theta (r) \dot{\theta} &=& f_\theta - \partial_\theta U(r, \theta,B) + 
\eta_\theta \circ \sqrt{2 k_B T \Gamma_\theta (r)}, \nonumber 
\eea
where $\circ$ denotes the Stratonovich product, $\Gamma_r(r)$ and $\Gamma_\theta(r)$ denote
friction coefficients, and $\eta_i$ is a white noise with $i \equiv \{ r, \theta \}$ such that
$\langle \eta_i (t) \eta_j (t') \rangle = \delta_{ij} \delta(t-t')$.
Drift terms $f_r=- \partial_r \ln \Gamma_r (r)/2$ and $f_\theta=- \partial_\theta \ln \Gamma_\theta (r)/2$ are chosen
such that the dynamics converge towards equilibrium for a constant magnetic field \cite{Lau2007_vol76,VanKampen2007_vol}. 

Dissipation in this system is mainly of hydrodynamic origin. In view of the proximity of the two beads 
with respect to each other and to the wall, one can rely on the lubrication approximation to describe
the hydrodynamic friction coefficients. 
These coefficients are the sum of the friction due to the sphere-sphere interaction $\Gamma_i^{s}$ and the friction
between the sphere and the bottom wall $\Gamma_i^{w}$. More explicitly $\Gamma_i(r)=
 \Gamma_i^{s} + \Gamma_i^{w}$ for $i \equiv \{ r, \theta \}$, with
\bea
\label{friction}
\Gamma_r^{s} (r) &=& \frac{\gamma a}{4 r -8 a}, \,\,\,
 \Gamma_r^{w} (r)= \frac{8}{15} \gamma \ln \frac{a}{b},  \\ 
\Gamma_\theta^{s} (r) &=& \frac{\gamma r^2}{2 k(r)}, \,\,\,  
\Gamma_\theta^{w} (r) = \frac{8}{15} \gamma r^2 \ln \frac{a}{b}, \nonumber
\eea
where $a$ is the bead radius, $b$ is the distance between the beads and the wall,
$\gamma$ is the bare friction coefficient of a single bead far from the wall and $k(r)$ is
a function given in Supl. Mat. and in Ref.~\cite{Jeffrey1984_vol139}. 

\begin{figure}[t!]
\includegraphics[scale=0.3]{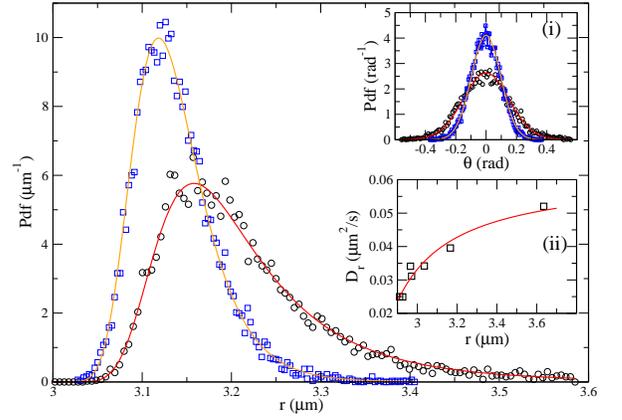}
\hspace{2mm}
\caption{\label{fig2}
Probability distribution function (pdf) of the relative distance between the beads 
for two values of the magnetic field, namely $B=0.3$mT (circles) and $B=0.45$mT (squares).
In the inset (i), the pdf of the angle $\theta$ is shown
for these two magnetic fields with corresponding symbols and in (ii), the
measurements of the radial diffusion coefficient $D_r(r)$ (symbols) are shown as function of   
the distance between the beads $r$ (in unit $\mu$m), together 
 with the theoretical prediction using Eq.~3 (solid line).}
\end{figure}

In order to test this model, we have measured experimentally the radial time auto-correlation function. 
The short time behavior of this function gives the radial diffusion coefficient $D_r(r)=k_B T/\Gamma_r(r)$. 
The data points can be well fitted to Eq.~\ref{friction} 
as shown in the inset (ii) of figure \ref{fig2}. From this fit, one finds that the diffusion coefficient 
of a single bead far from the wall is $D_0=k_B T/\gamma=0.12 \mu$m$^2$s$^{-1}$. 
This value is rather close to the Stokes-Einstein estimate $0.15 \mu$m$^2$s$^{-1}$ for  
a bead of diameter $2.805 \mu$m in water.

Within this framework, we study the distributions of thermodynamic quantities 
like work $W$ and heat $Q$, defined at the trajectory level by \cite{Sekimoto1998}
\bea
W(\tau) &=& \int_0^\tau dt \; \dot{B}(t) \partial_B U({\bf r}(t),B(t)), \\
Q(\tau) &=& \int_0^\tau dt \; \nabla_{\bf r} U({\bf r}(t),B(t)) \circ \dot{\bf r}. \nonumber 
\label{thermo_def}
\eea 
In order to make sure that the system 
is well equilibrated with a sufficient duration of the pauses, we have compared the equilibrium heat 
fluctuations $P_{eq}(Q)$ in a constant magnetic field with the internal energy fluctuations $P(\Delta U)$ evaluated
 in the out of equilibrium experiment, where $\Delta U$ represents the difference of internal energy between the end 
and the beginning of the cycle. If the system is well equilibrated, both distributions $P_{eq}(Q)$ and 
$P(\Delta U)$ should look identical as they do in fig. 3 of Suppl. Mat.

\begin{figure}[t!]
\includegraphics[width=\columnwidth]{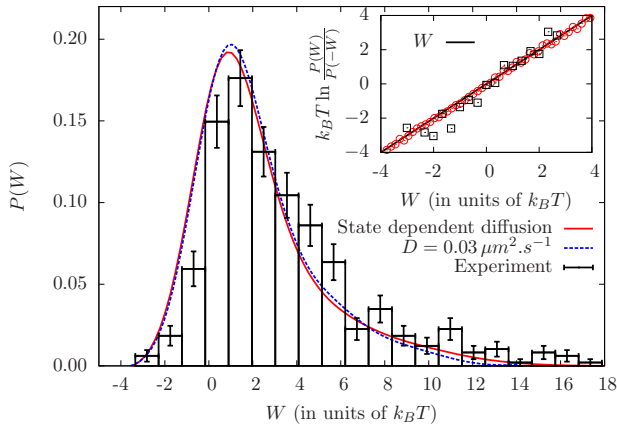} 
\hspace{2mm}
\caption{\label{fig3}
Probability distribution of the work $P(W)$, constructed from an experiment using 460 cycles (histogram) and 
from simulations of Eqs.~\ref{Langevin}-\ref{friction} with state-dependent diffusion coefficient (red solid line) and 
 with a constant diffusion coefficient $D=0.03\mu$m$^2$s$^{-1}$ (blue dashed line). 
In this experiment, $\tau=2$s, $\tau_{eq}=4$s and the sampling frequency is 40Hz.  
The inset shows the verification of the Crooks relation, namely the straight solid line, together with the 
experimental data (squares) and the simulation results (circles).}
\end{figure}

Using experimental trajectories corresponding to $\tau=2$s, we 
find an average work of $3.3 \pm 0.2 k_B T$ and a standard deviation of $3.6 k_B T$.
Note that $\langle W \rangle \geq 0$, as expected from the
 second law of thermodynamics which imposes 
that the dissipated work, $W_{diss}=W-\Delta F$ be on average positive. 
In the present case, 
$W_{diss}=W$ since the free energy difference $\Delta F=0$ for this symmetric
 protocol.
We have also evaluated the distribution of the work $P(W)$ represented in figure \ref{fig3},
which is non-Gaussian and agrees with the simulations of Eqs.~\ref{Langevin}. 
 We denote $\beta=1/k_B T$.
In the inset, we show that $P(W)$ satisfies the Crooks relation \cite{Crooks2000_vol61}
\beq
\ln \frac{P(W)}{P(-W)} = \beta W,
\label{crooks}
\eeq
both for the experimental data and for the simulations.
We observe that the relation holds in a smaller range for the experimental data than
for the simulations data due to a lack of statistics in the experiment 
($460$ trajectories in the particular experiment of figure \ref{fig3}). 

In this figure, we also compare simulations of the work distributions with the state-dependent diffusion coefficient 
given by Eq.~\ref{friction} and with a constant diffusion coefficient  $D=0.03\mu$m$^2$s$^{-1}$, corresponding 
to a typical distance between the beads. 
This comparison shows that the distribution of work is only weakly 
sensitive to the space-dependence of the diffusion coefficient. We attribute this
to the small excursion experienced by the particles in the cycle. In principle, if this
excursion could be made larger while still maintaining a sufficient confinement for a cyclic
operation to be possible, one may observe a stronger impact of the space-dependence of the friction coefficient 
on thermodynamic quantities.
In any case, the present study of stochastic thermodynamics in the presence of a space-dependent friction represents 
our first result. 

Until now, $\langle W \rangle$ could only be evaluated from the interaction potential, and such a determination 
had to be very accurate because a relatively small 
change of the control parameter can produce large variations of this potential. 
In view of this limitation, it would be interesting to develop alternative 
methods to estimate the average work, which would ideally bypass the need of a precise calibration of the potential. 
Since $\langle W \rangle$ is simply related to the average dissipation (the
system is in contact with a single thermostat), what we need in fact is an 
estimate of the average dissipation using only trajectory information.
On the theoretical side, it has been established that 
$\langle W_{diss} \rangle$ is related to the determination of an arrow of time,
by C. Jarzynski \cite{Jarzynski2006} and by R. Kawai et al.  
  \cite{Kawai2007_vol98} for hamiltonian systems, and by G. Crooks \cite{Crooks2000_vol61} and 
P. Gaspard \cite{Gaspard2004_vol117a} for systems in contact with a reservoir. 
In this latter case and for a reservoir of temperature $T$, one has the following 
 equality,
\beq
\beta \langle W_{diss} \rangle = D(P_F[c^F] || P_R[c^R]),
\label{Kawai form}
\eeq
where $D(P_F || P_R)$ represents 
the Kullback-Leibler (KL) divergence between the forward path  
probability $P_F$ evaluated on the forward trajectory $[c^F]$ 
and the corresponding probability
 distribution $P_R$ evaluated on the backward trajectory $[c^R]$.

In contrast to early studies on fluctuations theorem, which ``verified'' a relation like Eq.~\ref{crooks} from a determination of $P(W)$ as done in Fig.~\ref{fig3}, 
the above result suggests to do just the opposite: namely enforce the fluctuation relation 
as a constraint and estimate from it the dissipation using trajectories information.
This is indeed possible as shown in \cite{Andrieux2008a} via a careful analysis of 
the continuous times series 
of non-equilibrium stationary fluctuations. 
For discrete time series, a similar idea was put forward in 
\cite{Roldan2010_vol105} and recently used by two of us 
for estimating in a non-invasive way the dissipation present in chemical reactions \cite{Muy2013}.
So far, these ideas have not been exploited experimentally for driven non-stationary systems or for colloids in 
non-harmonic potentials.

\begin{figure}
\vspace{0.6cm}
\includegraphics[scale=0.35, trim= 3 3 3 3, clip=true]{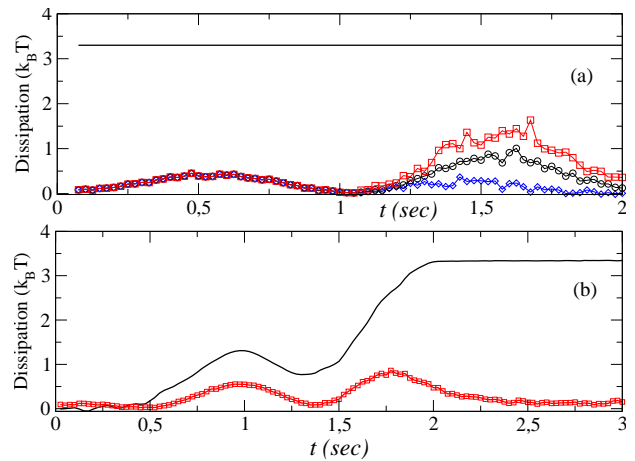}
\caption{\label{fig4}
a): Energy-based estimate of dissipation expressed in units of $k_B T$ 
(horizontal solid line) compared with an information-based estimate based on Eq. \ref{new Kawai form} 
(symbols) versus time $t$. The three estimators 
correspond to discarding the points where a log-divergence occurs (blue diamonds), 
using an adapted binning (black circles), or adding a small bias equal to $1/460$ 
to prevent log-divergences (red squares). 
b): Energy-based estimate of the dissipated work (black solid line) 
compared with an information-based estimate 
based on Eq. \ref{KL_bound} (red squares). 
}
\end{figure}

In order to do so, we now project the path probabilities $P_F[c^F]$ onto the probability distribution evaluated at a single point at time $t$ in the trajectory, namely $p_F(t)$. Similarly, $P_R[c^R]$ is projected onto the distribution evaluated at the time-symmetric point $p_R(\tau-t)$. As a result of these projections, Eq.~\ref{Kawai form} becomes the following inequality \cite{Horowitz2009}, which holds for any times $t$ and $\tau$:
\beq
\beta \langle W_{diss}(\tau) \rangle \geq D(p_F(t) || p_R(\tau-t)).
\label{new Kawai form}
\eeq
Now, taking advantage of the symmetry of the protocol, we can evaluate $p_R$ from the forward protocol. 
In other words, we record the trajectory at only two points in the cycle: the first one at a time 
$t$ after the beginning of the cycle and the second one at a time $\tau-t$ as shown in fig.~\ref{fig1}. 
We use only the information contained in the relative distance between the beads $r$ instead of $(r,\theta)$ since
our numerical simulations indicate that reliable estimates of the work can already be obtained in this way.
The probability distribution is determined from the experimental data after binning the trajectories, 
and from these the KL bound is evaluated as shown in fig.~\ref{fig4}a.
Note that by construction the bound is zero at $t=1$s 
where both measurement points merge into a single point. 
More interestingly, there is a maximum in this bound which occurs roughly half-way 
through the second half of the protocol at a time $t=1.6$s, 
and at this point a value of about 1.5 $k_B T$ is obtained. 
The precise value of the average dissipation depends on how the KL divergence is evaluated.
While all the estimators agree with each other when $t \leq 1$s, a notable difference  
between them is present near the maximum at
$t \simeq 1.6$s, where a log divergence occurs in the data due to lack of statistics.
The simplest strategy, namely to discard the points where a log-divergence occurs, gives the lowest estimate of the KL divergence. 
Alternatively, one can either bin the data in such a way that these divergences do not occur, or introduce a small constant bias in the probabilities equal to $1/460$ in order to remove the divergences.  
Both methods lead consistently to a higher value, in the range of 1 to 1.5 $k_B T$ near this maximum.
It is also important to appreciate that for all times $t$ considered, 
the KL bound is always smaller than the value obtained from the ``energetic'' estimate using the potential, which gives
the constant $3.3 \pm 0.2 \, k_B T$, independent of $t$. This is expected since most of the 
information contained in the trajectories has been discarded in the projection step to obtain 
Eq.~\ref{new Kawai form}, and only the values at two symmetric points were kept. This 
loss of information represents a form of coarse-graining which is known to lead to an underestimation
 of the dissipation 
\cite{Kawai2007_vol98},\cite{Horowitz2009},\cite{Esposito2012_vol85}. 

Since an equilibrium probability distribution is typically better known   
experimentally than its non-equilibrium counterpart, one may be tempted to replace the above comparison 
between forward and backward non-equilibrium probabilities, by a comparison 
between a non-equilibrium probability distribution, $p_{neq}(t)$, 
with its equilibrium counterpart, $p_{eq}(t)$. In such a formulation, 
the equilibrium distribution must be evaluated at the value of the control parameter at time $t$ 
and compared with the non-equilibrium distribution at the same time $t$ according to
\cite{Vaikuntanathan2009_vol87}:
\beq
\beta \langle W_{diss}(t) \rangle \geq D(p_{neq}(t) || p_{eq}(t)),
\label{KL_bound}
\eeq
where $\langle W_{diss}(t) \rangle$ is the average dissipative work evaluated up to time $t$. 
Note that for our specific experimental conditions, Eq. \ref{KL_bound} 
is only a particular case of Eq.~\ref{new Kawai form} when the time $t=\tau$.
In fig.~\ref{fig4}b, both sides of the inequality of Eq.~\ref{KL_bound} are evaluated 
for the same experimental data used in figure 4a as explained in more details in Supl. Mat. 
At large time $t$, $\langle W_{diss}(t) \rangle$ tends towards the average work determined before.
At short time $t$, both $\langle W_{diss}(t) \rangle$ and $D(p_{neq}(t) || p_{eq}(t))$ should tend to 
zero, but a small non-zero value is found in the latter quantity. We attribute 
this discrepancy to a small error in the determination of
the interaction potential which enters in $p_{eq}$. 
More importantly, the KL bound reaches a maximum of the order of 1$k_B T$
at a time $t \simeq 1.8$s, so somewhere within the second half of the cycle. Therefore, the 
amplitude of the estimated dissipation and its location in time are both consistent with the determination
 based on Eq. \ref{new Kawai form}.

To summarize, we have performed a test of stochastic thermodynamics in a system with a space-dependent friction, 
a friction which we have measured experimentally and characterized with an hydrodynamic model. 
In a first step, we have followed an energetic approach based on the determination of an interaction potential. 
In many complex systems, this energy-based approach is not practical, because the precise determination of the potential is too cumbersome or simply because there are too many variables
involved. To address this fundamental issue, we have investigated in a second step, 
information-theoretic estimations of the average dissipation. 
Of particular interest is the general formulation based on Eq.~\ref{Kawai form}, 
which has the advantage of not requiring any knowledge of the energetics of the system or of its equilibrium behavior. 
Both estimates are lower than the expected level of dissipation, and to improve upon this, 
extensions of this method are needed to take advantage of the complete 
information contained in the trajectories as opposed to only the information in a few points as done here.
Despite this limitation, information-theoretic estimates are attractive since they are
 simple to implement and 
  do not require any knowledge of the dynamics of the system, a definitive advantage 
  for many experimental applications.
In particular, we envision that this method could be useful for the monitoring of small chemical or 
biochemical reactors \cite{Muy2013} or for microrheology studies of biological systems.

We acknowledge insightful discussions with H. Stone, J. M. R. Parrondo and M. Esposito.
D. L. would also like to thank the Kavli Institute for Theoretical Physics China, CAS, Beijing 100190, 
China, for hospitality in the summer of 2013, during which part of this work was done.

%
\bibliographystyle{myapsrev}
\bibliography{anupam}

\begin{thebibliography}{23}
\expandafter\ifx\csname natexlab\endcsname\relax\def\natexlab#1{#1}\fi
\expandafter\ifx\csname bibnamefont\endcsname\relax
  \def\bibnamefont#1{#1}\fi
\expandafter\ifx\csname bibfnamefont\endcsname\relax
  \def\bibfnamefont#1{#1}\fi
\expandafter\ifx\csname citenamefont\endcsname\relax
  \def\citenamefont#1{#1}\fi
\expandafter\ifx\csname url\endcsname\relax
  \def\url#1{\texttt{#1}}\fi
\expandafter\ifx\csname urlprefix\endcsname\relax\def\urlprefix{URL }\fi
\providecommand{\bibinfo}[2]{#2}
\providecommand{\eprint}[2][]{\url{#2}}

\bibitem[{\citenamefont{{Seifert}}(2012)}]{Seifert2012}
\bibinfo{author}{\bibfnamefont{U.}~\bibnamefont{{Seifert}}},
  \bibinfo{journal}{Rep. Prog. Phys.} \textbf{\bibinfo{volume}{75}},
  \bibinfo{pages}{126001} (\bibinfo{year}{2012}).

\bibitem[{\citenamefont{Jarzynski}(2011)}]{Jarzynski2011_vol2}
\bibinfo{author}{\bibfnamefont{C.}~\bibnamefont{Jarzynski}},
  \bibinfo{journal}{Annu. Rev. Condens. Mat. Phys.}
  \textbf{\bibinfo{volume}{2}}, \bibinfo{pages}{329} (\bibinfo{year}{2011}).

\bibitem[{\citenamefont{Ciliberto et~al.}(2010)\citenamefont{Ciliberto,
  Joubaud, and Petrosyan}}]{Ciliberto2010_vol}
\bibinfo{author}{\bibfnamefont{S.}~\bibnamefont{Ciliberto}},
  \bibinfo{author}{\bibfnamefont{S.}~\bibnamefont{Joubaud}}, \bibnamefont{and}
  \bibinfo{author}{\bibfnamefont{A.}~\bibnamefont{Petrosyan}},
  \bibinfo{journal}{J. Stat. Mech.}  \bibinfo{pages}{P12003}
  (\bibinfo{year}{2010}).

\bibitem[{\citenamefont{Ritort}(2008)}]{Ritort2008_vol137}
\bibinfo{author}{\bibfnamefont{F.}~\bibnamefont{Ritort}},
  \bibinfo{journal}{Adv. Chem. Phys.} \textbf{\bibinfo{volume}{137}},
  \bibinfo{pages}{31} (\bibinfo{year}{2008}).

\bibitem[{\citenamefont{Imparato et~al.}(2007)\citenamefont{Imparato, Peliti,
  Pesce, Rusciano, and Sasso}}]{Imparato2007_vol76}
\bibinfo{author}{\bibfnamefont{A.}~\bibnamefont{Imparato}},
  \bibinfo{author}{\bibfnamefont{L.}~\bibnamefont{Peliti}},
  \bibinfo{author}{\bibfnamefont{G.}~\bibnamefont{Pesce}},
  \bibinfo{author}{\bibfnamefont{G.}~\bibnamefont{Rusciano}}, \bibnamefont{and}
  \bibinfo{author}{\bibfnamefont{A.}~\bibnamefont{Sasso}},
  \bibinfo{journal}{Phys. Rev. E} \textbf{\bibinfo{volume}{76}},
  \bibinfo{pages}{050101} (\bibinfo{year}{2007}).

\bibitem[{\citenamefont{Blickle et~al.}(2006)\citenamefont{Blickle, Speck,
  Helden, Seifert, and Bechinger}}]{Blickle2006_vol96}
\bibinfo{author}{\bibfnamefont{V.}~\bibnamefont{Blickle}},
  \bibinfo{author}{\bibfnamefont{T.}~\bibnamefont{Speck}},
  \bibinfo{author}{\bibfnamefont{L.}~\bibnamefont{Helden}},
  \bibinfo{author}{\bibfnamefont{U.}~\bibnamefont{Seifert}}, \bibnamefont{and}
  \bibinfo{author}{\bibfnamefont{C.}~\bibnamefont{Bechinger}},
  \bibinfo{journal}{Phys. Rev. Lett.} \textbf{\bibinfo{volume}{96}},
  \bibinfo{pages}{070603} (\bibinfo{year}{2006}).

\bibitem[{\citenamefont{Celani et~al.}(2012)\citenamefont{Celani, Bo, Eichhorn,
  and Aurell}}]{Celani2012}
\bibinfo{author}{\bibfnamefont{A.}~\bibnamefont{Celani}},
  \bibinfo{author}{\bibfnamefont{S.}~\bibnamefont{Bo}},
  \bibinfo{author}{\bibfnamefont{R.}~\bibnamefont{Eichhorn}}, \bibnamefont{and}
  \bibinfo{author}{\bibfnamefont{E.}~\bibnamefont{Aurell}},
  \bibinfo{journal}{Phys. Rev. Lett.} \textbf{\bibinfo{volume}{109}},
  \bibinfo{pages}{260603} (\bibinfo{year}{2012}).

\bibitem[{\citenamefont{Brangbour et~al.}(2011)\citenamefont{Brangbour,
  du~Roure, Helfer, D{\'e}moulin, Mazurier, Fermigier, Carlier, Bibette, and
  Baudry}}]{Brangbour2011}
\bibinfo{author}{\bibfnamefont{C.}~\bibnamefont{Brangbour}},
  \bibinfo{author}{\bibfnamefont{O.}~\bibnamefont{du~Roure}},
  \bibinfo{author}{\bibfnamefont{E.}~\bibnamefont{Helfer}},
  \bibinfo{author}{\bibfnamefont{D.}~\bibnamefont{D{\'e}moulin}},
  \bibinfo{author}{\bibfnamefont{A.}~\bibnamefont{Mazurier}},
  \bibinfo{author}{\bibfnamefont{M.}~\bibnamefont{Fermigier}},
  \bibinfo{author}{\bibfnamefont{M.-F.} \bibnamefont{Carlier}},
  \bibinfo{author}{\bibfnamefont{J.}~\bibnamefont{Bibette}}, \bibnamefont{and}
  \bibinfo{author}{\bibfnamefont{J.}~\bibnamefont{Baudry}},
  \bibinfo{journal}{PLoS Biol} \textbf{\bibinfo{volume}{9}},
  \bibinfo{pages}{e1000613} (\bibinfo{year}{2011}).

\bibitem[{\citenamefont{Lacoste et~al.}(2009)\citenamefont{Lacoste, Brangbour,
  Bibette, and Baudry}}]{Lacoste2009_vol80}
\bibinfo{author}{\bibfnamefont{D.}~\bibnamefont{Lacoste}},
  \bibinfo{author}{\bibfnamefont{C.}~\bibnamefont{Brangbour}},
  \bibinfo{author}{\bibfnamefont{J.}~\bibnamefont{Bibette}}, \bibnamefont{and}
  \bibinfo{author}{\bibfnamefont{J.}~\bibnamefont{Baudry}},
  \bibinfo{journal}{Phys. Rev. E} \textbf{\bibinfo{volume}{80}},
  \bibinfo{pages}{011401} (\bibinfo{year}{2009}).

\bibitem[{\citenamefont{Lau and Lubensky}(2007)}]{Lau2007_vol76}
\bibinfo{author}{\bibfnamefont{A.~W.~C.} \bibnamefont{Lau}} \bibnamefont{and}
  \bibinfo{author}{\bibfnamefont{T.~C.} \bibnamefont{Lubensky}},
  \bibinfo{journal}{Phys. Rev. E} \textbf{\bibinfo{volume}{76}},
  \bibinfo{pages}{011123} (\bibinfo{year}{2007}).

\bibitem[{\citenamefont{Van~Kampen}(2007)}]{VanKampen2007_vol}
\bibinfo{author}{\bibfnamefont{N.}~\bibnamefont{Van~Kampen}},
  \emph{\bibinfo{title}{Stochastic Processes in Physics and Chemistry}}
  (\bibinfo{publisher}{North-Holland Personal Library}, \bibinfo{year}{2007}).

\bibitem[{\citenamefont{Jeffrey and Onishi}(1984)}]{Jeffrey1984_vol139}
\bibinfo{author}{\bibfnamefont{D.~J.} \bibnamefont{Jeffrey}} \bibnamefont{and}
  \bibinfo{author}{\bibfnamefont{Y.}~\bibnamefont{Onishi}},
  \bibinfo{journal}{J. Fluid Mech.} \textbf{\bibinfo{volume}{139}},
  \bibinfo{pages}{261} (\bibinfo{year}{1984}).

\bibitem[{\citenamefont{Sekimoto}(1998)}]{Sekimoto1998}
\bibinfo{author}{\bibfnamefont{K.}~\bibnamefont{Sekimoto}},
  \bibinfo{journal}{Prog. of Theo. Phys.} \textbf{\bibinfo{volume}{130}},
  \bibinfo{pages}{17} (\bibinfo{year}{1998}).

\bibitem[{\citenamefont{Crooks}(2000)}]{Crooks2000_vol61}
\bibinfo{author}{\bibfnamefont{G.~E.} \bibnamefont{Crooks}},
  \bibinfo{journal}{Phys. Rev. E} \textbf{\bibinfo{volume}{61}},
  \bibinfo{pages}{2361} (\bibinfo{year}{2000}).

\bibitem[{\citenamefont{Jarzynski}(2006)}]{Jarzynski2006}
\bibinfo{author}{\bibfnamefont{C.}~\bibnamefont{Jarzynski}},
  \bibinfo{journal}{Phys. Rev. E} \textbf{\bibinfo{volume}{73}},
  \bibinfo{pages}{046105} (\bibinfo{year}{2006}).

\bibitem[{\citenamefont{Kawai et~al.}(2007)\citenamefont{Kawai, Parrondo, and
  Van~den Broeck}}]{Kawai2007_vol98}
\bibinfo{author}{\bibfnamefont{R.}~\bibnamefont{Kawai}},
  \bibinfo{author}{\bibfnamefont{J.~M.~R.} \bibnamefont{Parrondo}},
  \bibnamefont{and} \bibinfo{author}{\bibfnamefont{C.}~\bibnamefont{Van~den
  Broeck}}, \bibinfo{journal}{Phys. Rev. Lett.} \textbf{\bibinfo{volume}{98}},
  \bibinfo{pages}{080602} (\bibinfo{year}{2007}).

\bibitem[{\citenamefont{Gaspard}(2004)}]{Gaspard2004_vol117a}
\bibinfo{author}{\bibfnamefont{P.}~\bibnamefont{Gaspard}}, \bibinfo{journal}{J.
  Stat. Phys.} \textbf{\bibinfo{volume}{117}}, \bibinfo{pages}{599}
  (\bibinfo{year}{2004}).

\bibitem[{\citenamefont{Andrieux et~al.}(2008)\citenamefont{Andrieux, Gaspard,
  Ciliberto, Garnier, Joubaud, and Petrosyan}}]{Andrieux2008a}
\bibinfo{author}{\bibfnamefont{D.}~\bibnamefont{Andrieux}},
  \bibinfo{author}{\bibfnamefont{P.}~\bibnamefont{Gaspard}},
  \bibinfo{author}{\bibfnamefont{S.}~\bibnamefont{Ciliberto}},
  \bibinfo{author}{\bibfnamefont{N.}~\bibnamefont{Garnier}},
  \bibinfo{author}{\bibfnamefont{S.}~\bibnamefont{Joubaud}}, \bibnamefont{and}
  \bibinfo{author}{\bibfnamefont{A.}~\bibnamefont{Petrosyan}},
  \bibinfo{journal}{J. Stat. Mech.}  \bibinfo{pages}{P01002}
  (\bibinfo{year}{2008}).

\bibitem[{\citenamefont{{Rold\'an} and Parrondo}(2010)}]{Roldan2010_vol105}
\bibinfo{author}{\bibfnamefont{{\'E}.}~\bibnamefont{{Rold\'an}}}
  \bibnamefont{and} \bibinfo{author}{\bibfnamefont{J.~M.~R.}
  \bibnamefont{Parrondo}}, \bibinfo{journal}{Phys. Rev. Lett.}
  \textbf{\bibinfo{volume}{105}}, \bibinfo{pages}{150607}
  (\bibinfo{year}{2010}).

\bibitem[{\citenamefont{Muy et~al.}(2013)\citenamefont{Muy, Kundu, and
  Lacoste}}]{Muy2013}
\bibinfo{author}{\bibfnamefont{S.}~\bibnamefont{Muy}},
  \bibinfo{author}{\bibfnamefont{A.}~\bibnamefont{Kundu}}, \bibnamefont{and}
  \bibinfo{author}{\bibfnamefont{D.}~\bibnamefont{Lacoste}},
  \bibinfo{journal}{J. Chem. Phys.} \textbf{\bibinfo{volume}{139}},
  \bibinfo{pages}{124109} (\bibinfo{year}{2013}).

\bibitem[{\citenamefont{Horowitz and Jarzynski}(2009)}]{Horowitz2009}
\bibinfo{author}{\bibfnamefont{J.}~\bibnamefont{Horowitz}} \bibnamefont{and}
  \bibinfo{author}{\bibfnamefont{C.}~\bibnamefont{Jarzynski}},
  \bibinfo{journal}{Phys. Rev. E} \textbf{\bibinfo{volume}{79}},
  \bibinfo{pages}{021106} (\bibinfo{year}{2009}).

\bibitem[{\citenamefont{Esposito}(2012)}]{Esposito2012_vol85}
\bibinfo{author}{\bibfnamefont{M.}~\bibnamefont{Esposito}},
  \bibinfo{journal}{Phys. Rev. E} \textbf{\bibinfo{volume}{85}},
  \bibinfo{pages}{041125} (\bibinfo{year}{2012}).

\bibitem[{\citenamefont{Vaikuntanathan and
  Jarzynski}(2009)}]{Vaikuntanathan2009_vol87}
\bibinfo{author}{\bibfnamefont{S.}~\bibnamefont{Vaikuntanathan}}
  \bibnamefont{and}
  \bibinfo{author}{\bibfnamefont{C.}~\bibnamefont{Jarzynski}},
  \bibinfo{journal}{Europhys. Lett.} \textbf{\bibinfo{volume}{87}},
  \bibinfo{pages}{60005} (\bibinfo{year}{2009}).

\end{thebibliography}


\begin{thebibliography}{4}
\expandafter\ifx\csname natexlab\endcsname\relax\def\natexlab#1{#1}\fi
\expandafter\ifx\csname bibnamefont\endcsname\relax
  \def\bibnamefont#1{#1}\fi
\expandafter\ifx\csname bibfnamefont\endcsname\relax
  \def\bibfnamefont#1{#1}\fi
\expandafter\ifx\csname citenamefont\endcsname\relax
  \def\citenamefont#1{#1}\fi
\expandafter\ifx\csname url\endcsname\relax
  \def\url#1{\texttt{#1}}\fi
\expandafter\ifx\csname urlprefix\endcsname\relax\def\urlprefix{URL }\fi
\providecommand{\bibinfo}[2]{#2}
\providecommand{\eprint}[2][]{\url{#2}}

\bibitem[{\citenamefont{Lacoste et~al.}(2009)\citenamefont{Lacoste, Brangbour,
  Bibette, and Baudry}}]{Lacoste2009_vol80}
\bibinfo{author}{\bibfnamefont{D.}~\bibnamefont{Lacoste}},
  \bibinfo{author}{\bibfnamefont{C.}~\bibnamefont{Brangbour}},
  \bibinfo{author}{\bibfnamefont{J.}~\bibnamefont{Bibette}}, \bibnamefont{and}
  \bibinfo{author}{\bibfnamefont{J.}~\bibnamefont{Baudry}},
  \bibinfo{journal}{Phys. Rev. E} \textbf{\bibinfo{volume}{80}},
  \bibinfo{pages}{011401} (\bibinfo{year}{2009}).

\bibitem[{\citenamefont{Russel et~al.}(1989)\citenamefont{Russel, Saville, and
  Schowalter}}]{russel}
\bibinfo{author}{\bibfnamefont{W.~B.} \bibnamefont{Russel}},
  \bibinfo{author}{\bibfnamefont{D.}~\bibnamefont{Saville}}, \bibnamefont{and}
  \bibinfo{author}{\bibfnamefont{W.~R.} \bibnamefont{Schowalter}},
  \emph{\bibinfo{title}{Colloidal Dispersions}} (\bibinfo{publisher}{Cambridge
  University Press}, \bibinfo{address}{Cambridge}, \bibinfo{year}{1989}).

\bibitem[{\citenamefont{Jeffrey and Onishi}(1984)}]{Jeffrey1984_vol139}
\bibinfo{author}{\bibfnamefont{D.~J.} \bibnamefont{Jeffrey}} \bibnamefont{and}
  \bibinfo{author}{\bibfnamefont{Y.}~\bibnamefont{Onishi}},
  \bibinfo{journal}{J. Fluid Mech.} \textbf{\bibinfo{volume}{139}},
  \bibinfo{pages}{261} (\bibinfo{year}{1984}).

\bibitem[{Note1()}]{Note1}
Note1, \bibinfo{note}{we keep the same notation as Jeffrey and Onishi but in
  the following of the paper, the letter $x$ and $y$ will refer to the
  cartesian coordinates of the fictive particle}.

\end{thebibliography}

\end{document}


%
%

\title {Supplementary information for 'Energy versus information based estimations of dissipation using a pair of magnetic colloidal particles'}
\author{S. Tusch$^2$, A. Kundu$^1$, G. Verley$^1$, T. Blondel$^1$, V. Miralles$^2$, 
D. D{\'e}moulin$^2$, D. Lacoste$^1$, J. Baudry$^2$}
\affiliation{$^1$ Laboratoire de Physico-Chimie Th\'eorique - UMR CNRS Gulliver 7083,
ESPCI, 10 rue Vauquelin, F-75231 Paris, France}
\affiliation{$^2$ Laboratoire LCMD, ESPCI, 10 rue Vauquelin, F-75231 Paris, France}
\date{\today}

\maketitle

\blfootnote{A. Kundu is presently employed at Laboratoire de Physique Th\'eorique et Mod\`edes Statistiques - UMR CNRS 8626,
Universit\'e Paris-Sud, B\^at. 100, 91405 Orsay Cedex, France. G. Verley is employed at Physics and Material Sciences Research Unit, 
University of Luxembourg, L-1511 Luxembourg, G.D. Luxembourg.}

\section{Interaction potential between the beads}
We use the 2D relative displacement vector in polar coordinates ${\bf r} =(r,\theta)$ as shown in (fig1).
The interaction between the beads is modeled using a potential, 
which is the sum of three contributions: the dipolar interaction of the magnetic beads with each 
other $U_{dip}$, the interaction $U_{mag}$ of the beads with the applied magnetic 
field ${\bf B}=B \hat z$, 
and a repulsive interaction of electrostatic origin $U_{el}$: 
\beq
U(r,\theta,B)=U_{dip}(B,r,\theta) + U_{mag}(B) + U_{el}(r).
\eeq
This potential has a short range repulsive part due to the electrostatics and
a long-range attractive part due to the dipolar interaction as described
 in \cite{Lacoste2009_vol80}. It is expressed in units of $k_B T$.
We provide below the values of the experimental parameters entering in this potential 
corresponding to the $\tau=2$s experiment
described in the main text.

The electrostatic part of the potential is obtained from Debye-H\"uckel theory
adapted to the case of two spheres using the Derjaguin approximation. 
In the present case, where the particle diameter $d$ is much larger than 
the Debye length $\lambda_{DB}$, the expression is  
\beq
U_{el}(r)=U_0 \ln \left( 1 + e^{-(r-d)/\lambda_{DB}} \right), 
\eeq
where $U_0$ is the strength of the interaction, which depends on the particle electrostatic 
potential (zeta potential), the dielectric constant of the bead and the particle diameter. 
In the present experiment, we obtain from the fit of the probability distribution 
of the relative displacement between the beads: $U_0 \simeq 1800$ in units of $k_BT$,
$d=2.805 \mu$m, and $\lambda_{DB}=44.2$nm.

The magnetic dipolar part of the potential has the form
\beq
U_{dip}(B,r,\theta)=\left( \frac{B}{B_0} \right)^2 \left( \frac{d}{r} \right)^3 
\left( 1 - 3 \cos^2 \theta \right),
\eeq
where $B$ is the applied magnetic field and $B_0=0.09008$mT.
In this model, $U_{dip}$ describes the interaction between the two magnetic dipoles 
${\bf m}_1$ and ${\bf m}_2$ carried by the beads. 
As mentioned in the main text, since the value of the applied field is rather large 
(the corresponding energy is large with respect to $k_B T$), we can consider
that the orientation of these dipoles is frozen along the applied magnetic field.
Furthermore, we assume that these dipoles are independent of the bead distance and that the 
two beads are identical (in particular that they have the same radius) 
which implies that ${\bf m}_1={\bf m}_2=m \hat{z}$.
In the end, the constant $(B/B_0)^2$ represents in fact $m^2/ 4 \pi \mu_0$, 
given the linear relation between the magnetic dipoles and the applied magnetic field.

The direct interaction of the magnetic dipoles ${\bf m}_1$ and ${\bf m}_2$ carried 
by both beads with the magnetic field is
\beq
U_{mag}(B)=- {\bf m}_1 \cdot {\bf B} - {\bf m}_2 \cdot {\bf B},
\eeq
Given our assumption that ${\bf m}_1={\bf m}_2$ independent of $r$, this term
represents a contribution which is quadratic in the field but constant in terms of the 
$r$ dependence. 

\section{Hydrodynamic model of the friction between the beads}
Dissipation in this system is mainly of hydrodynamic origin. In view of the proximity of the two beads 
with respect to each other and to the wall, one can rely on the lubrication approximation to describe
the hydrodynamic friction coefficients. 
These coefficients are the sum of the friction due to the sphere-sphere interaction $\Gamma_i^{s}$ 
and the friction
between the sphere and the bottom wall $\Gamma_i^{w}$, because the corresponding forces 
are parallel to each other.
More explicitly $\Gamma_i(r)=
 \Gamma_i^{s} + \Gamma_i^{w}$ for $i \equiv \{ r, \theta \}$. We thank H. Stone for insightful
discussions concerning the proper modeling of these friction coefficients.

Let us first discuss the friction between the spheres and the wall, and to obtain that
we first need to know the interaction of a single sphere with a wall. 
This hydrodynamic interaction can be calculated within the lubrication approximation. 
We are mainly interested in the case of the translation of the sphere 
in a tangent plane parallel to the 
wall (assuming no rotation of the bead). In this case, the friction is increased 
by a factor $16 \pi \ln(a/b)/15$, where $a$ is the bead radius
and $b$ the gap between the sphere and the wall, with respect to the Stokes-Einstein friction 
of the same sphere in a bulk fluid \cite{russel}. 
It follows from this that the friction coefficient 
between the two beads and the wall along the $\er$ and $\et$ directions take the following form
\bea
\Gamma_r^{w} (r)= \frac{8}{15} \gamma \ln \frac{a}{b},  \label{friction sw} \\ 
\Gamma_\theta^{w} (r) = \frac{8}{15} \gamma r^2 \ln \frac{a}{b}, \nonumber
\eea
where $\gamma$ represents the bare friction coefficient of a single bead far from the wall.

In order to model the hydrodynamic interaction between the beads, we have built our
model on the work of Jeffrey and Onishi \cite{Jeffrey1984_vol139}.
The first step is to reduce the motion of the two beads to the motion of one fictive particle 
in the frame of the center of mass. 
We denote by $\dot {\bm r}^\sigma $ the velocity vector of the magnetic bead $\sigma$ 
and $ \dot {r}^\sigma_i$ its coordinate $i$. The force applied on the bead is 
$\bm F^{\sigma} = - \nabla^\sigma V(\bm r^\sigma - \bm r^\omega) = \nabla^\omega V(\bm r^\sigma
 - \bm r^\omega) = - \bm F^\omega $ and its coordinates are $F^\sigma_i$. 
The symbol $\nabla^\omega$ is for the gradient calculated with the coordinates of the bead $\omega$. 
In the following we neglect the rotation of the beads and the hydrodynamics torque. 
The force and the velocity are given by
\beq
\dot r^\sigma_i = M^{\sigma \omega}_{ij} F^\omega_j,
\eeq
thanks to the linearity of the Stokes equation at low Reynolds number \cite{russel}. By 
convention, all repeated indices are summed over.
The spheres being symmetric, the mobility tensor $M$ is symmetric in the exchange of the two beads, 
i.e. $M^{\sigma \omega}_{ij} = M^{ \omega \sigma}_{ij}$. The relative velocity of the two beads
 is given by the vector $\dot {\bm r} = \dot {\bm r}^1 - \dot {\bm r}^2$ and the coordinates of 
that vector verify
\bea
\dot r_i &=&  M^{1 1}_{ij} F^1_j + M^{1 2}_{ij} F^2_j - M^{21}_{ij} F^1_j - 
 M^{2 2}_{ij} F^2_j \nonumber \\
& =&  2  \left(M^{1 1}_{ij} - M^{12}_{ij} \right) F^1_j.
\eea
Using the unit vector $ \bm e_r = {\bf r}/r$ of coordinates $e_1=1$ and $e_2=0$ in the polar 
basis $(\bm e_r,\bm e_\theta)$, the mobility tensor takes the following form
\beq
M^{\sigma \omega}_{ij} = x^{\sigma \omega} e_i e_j + y^{\sigma \omega} (\delta_{ij} - e_i e_j)
\eeq
where the coefficients $x^{\sigma \omega}$ and $y^{\sigma \omega}$ are dependant of the distance
 $r$ between the two beads. The coordinates of the relative velocity become then
\beq
\dot r_i = 2 (x^{1 1} - x^{12} )  e_i e_j F^1_j + 2 (y^{11}-y^{12})  (\delta_{ij} - e_i e_j) F^1_j,
\eeq
or more explicitly because $\dot r_1= \dot r $ and $\dot r_2 = r \dot \theta$ in polar coordinates
 (we have for the relative velocity in the polar basis 
$\dot {\bf r} = \dot r \er + r \dot \theta \et $). 
\bea
\dot r &=& 2 (x^{1 1} - x^{12} ) F^1_1 = - 2 (x^{1 1} - x^{12} ) \partial_r V(r,\theta), \\
r \dot \theta &=& 2 (y^{11}-y^{12}) F^1_2 =-  \frac{2(y^{11}-y^{12})}{r} \partial_\theta V(r,\theta).
\eea
We have used the coefficients $x^{\sigma \omega}$ and $y^{\sigma \omega}$ 
introduced by D.J. Jeffrey and Y. Onishi \cite{Jeffrey1984_vol139} 
in the nearly touching sphere limit 
\footnote{We keep the same notation as Jeffrey and Onishi but in the following of the paper, 
the letter $x$ and $y$ will refer to the cartesian coordinates of the fictive particle}. 
In this limit, we have
\bea
x^{11} - x^{12} &=& \frac{4}{\gamma} \left ( \frac{r}{d}-1 \right ), \\
y^{11}-y^{12} &=&  \frac{0.402 (\ln \xi^{-1})^2+2.96\ln \xi^{-1}+5.09}
{ \gamma \left( (\ln \xi^{-1})^2+6.04\ln \xi^{-1}+6.33 \right)} , \label{eq:orthotensor}
\eea
where $\xi = 2(r-d)/d=(r-2a)/a$ and $\gamma = 6 \pi \eta d / 2=6 \pi \eta a$ 
the friction coefficient of one particle alone in the fluid of viscosity $\eta$. 
In order to obtain a compact notation, we introduce the notation $k(r)=\gamma (y^{11}-y^{12})$, with 
the $y^{11}-y^{12}$ given in the equation above. It follows that 
the friction coefficients for the sphere-sphere interaction take the form
\bea
\Gamma_r^{s} (r) &=& \frac{\gamma a}{4 r -8 a}, \\ \label{friction ss}
\Gamma_\theta^{s} (r) &=& \frac{\gamma r^2}{2 k(r)}. \nonumber
\eea
When these friction coefficients are combined with the friction coefficients for the sphere-wall 
interaction given in Eq.~\ref{friction sw}, 
the full friction tensor and the Langevin equations given in the main text are obtained.

\section{Distributions of heat and internal energy}
\begin{figure}[t!]
\includegraphics[width=\columnwidth]{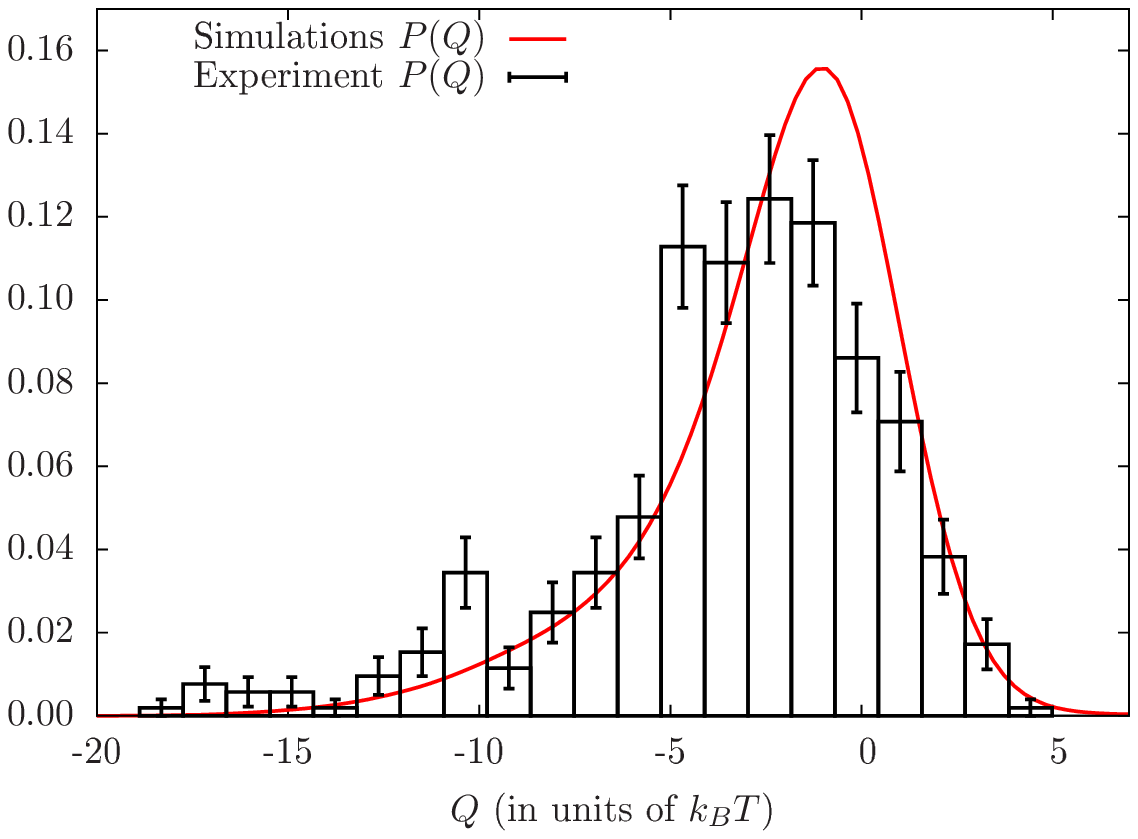}
\hspace{2mm}
\caption{\label{figS1}
Probability distribution of the heat $P(Q)$, constructed from an experiment using 460 cycles (histogram) and 
from a simulation of Eq.~2 of the main text (solid line), in the same conditions as the probability 
distribution of work shown in Fig~3 of the main text. 
The average heat in this experiment is $-3.4 \pm 0.2 k_B T$.}
\end{figure}

\begin{figure}[t!]
\includegraphics[width=\columnwidth]{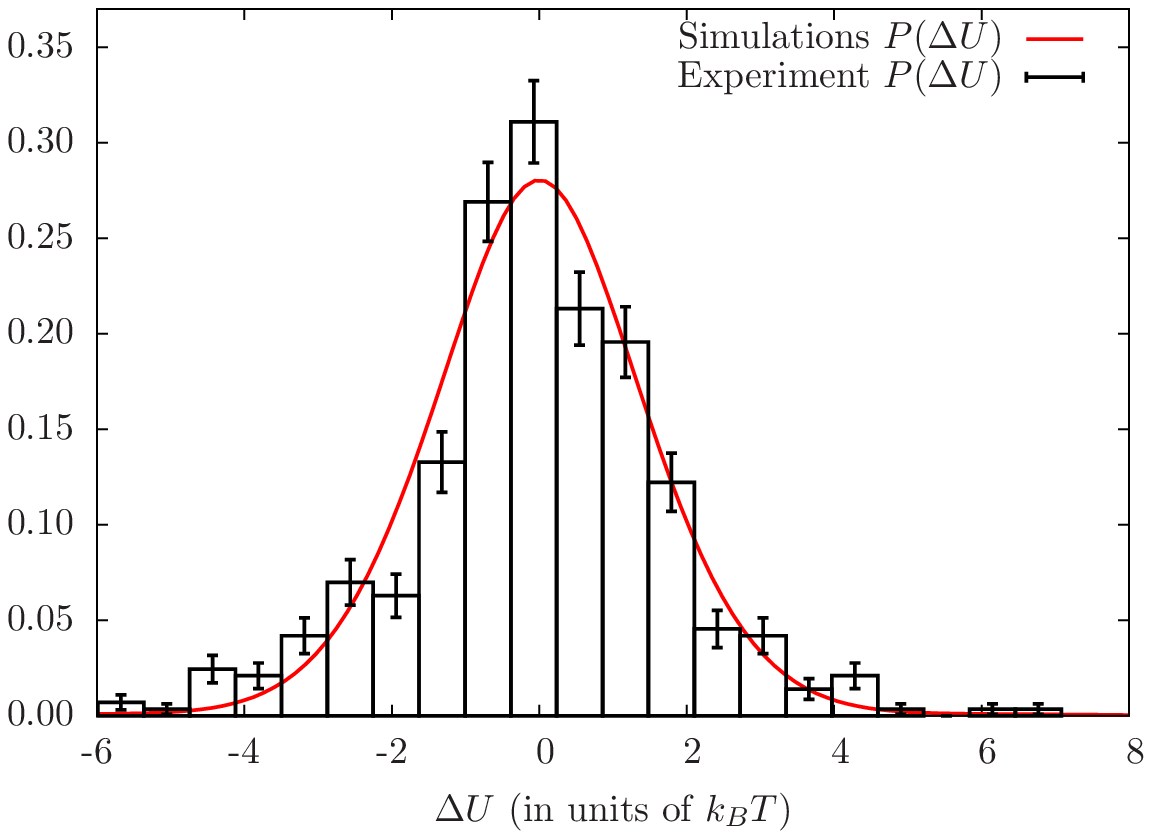}
\hspace{2mm}
\caption{\label{figS2}
Probability distribution of the internal energy $P(\Delta U)$, constructed from an experiment using 460 cycles (histogram) and 
from a simulation of Eq.~2 of the main text (solid line), in the same conditions as in Fig~1 of the main text. 
The average internal energy in this experiment is close to zero.}
\end{figure}

We show in figures \ref{figS1} and \ref{figS2} the probability distributions of heat and internal energy constructed from the experimental datas used also for figure 3 of the main text. 
These histograms represent experimental data points which 
are compared with simulations of the Langevin equations. 

\begin{figure}[t!]
\includegraphics[width=\columnwidth]{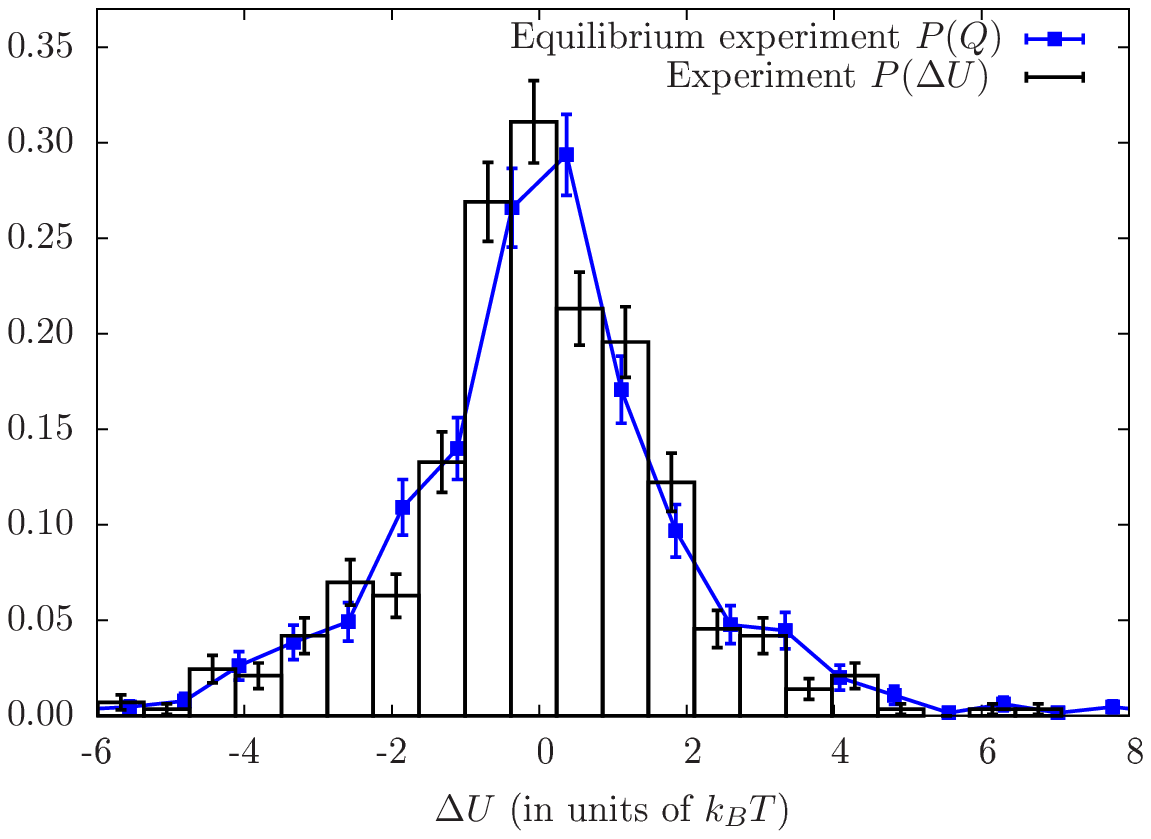}
\hspace{2mm}
\caption{\label{figS3}
Probability distribution of the internal energy change $P(\Delta U)$, constructed from non equilibrium 
experiments using 460 cycles (histogram) and probability distribution of heat exchange from equilibrium 
experiments (continuous line with data points).}
\end{figure}

With the definitions of work and heat of Eq.~4 of the main text, one can write the first law in the form of
$\Delta U= W +  Q$, where $\Delta U$ denotes the difference of internal energy 
between the initial and final point of the cycle namely, $U({\bf r}(\tau+\tau_{eq}),B(\tau+\tau_{eq}))-U({\bf r}(0),B(0))$.
Now, by definition of a cyclic protocol, the initial and final value of the control parameter
are the same. Since ${\bf r}(0)$ and ${\bf r}(\tau+\tau_{eq})$ are distributed according to the same canonical
distribution (if the equilibration is done properly), 
then it follows that $\langle \Delta U \rangle = \langle  W \rangle + \langle  Q \rangle=0$.
This is indeed well verified by the experimental average work and heat, since
we have obtained $\langle W \rangle=3.3 \pm 0.2 k_BT$ and $\langle Q \rangle =-3.4 \pm 0.2 k_B T$.
Naturally, higher moments of the distribution of the random variable $\Delta U$ may not vanish. 
In fact, as mentioned in the main text, a
strategy to test that the system is well equilibrated 
with a sufficient duration of the pauses consists precisely in studying this distribution of internal energy
and in comparing it with the heat fluctuations measured at equilibrium. 
Indeed, for a constant value of the magnetic field,
we measure the equilibrium distribution of heat, denoted $P_{eq}(Q)$.
Since there is no work in this case, this distribution $P_{eq}(Q)$ should match $P(\Delta U)$ in the non-equilibrium experiment 
provided the initial and end point in the cycle in the non-equilibrium experiment 
are well equilibrated and if the control parameter at these points has the same value as the one 
used in the equilibrium experiment.
Therefore, as mentioned in the text, a comparison between these two distributions offers a 
simple way to test equilibration in this system. As shown in fig~\ref{figS3}, this
test is well satisfied in the conditions of our experiment.

We have also studied the approach to the quasi-static limit using simulations.
Using simulations, we have evaluated $\langle W \rangle$ for different durations $\tau$ 
of the protocol. As expected $\langle W \rangle \to 0$ in the quasi-static limit $\tau \to \infty$
and in the limit $\tau \to 0$ by definition. As a result, we find that $\langle W \rangle$ has a 
maximum at some $\tau$ which depends on equilibration time $\tau_{eq}$ and on the relaxation time
$\tau_{rel}$ characteristic of the fluctuations around the minimum of the potential.
Since the potential is very anharmonic, $\tau_{eq} \gg \tau_{rel}$.
We have also observed that the maximum of $\langle W \rangle$ occurs at a time which is too short to be
accessible with our experiment.

A few words on how the experimental histograms have been constructed:
The histograms have been obtained by counting the number of events $n_i$ with 
a work, heat or energy change in the range of the $i$th 
bar of the histogram, i.e. $p_i = n_i/N$ with $N$ the total number of events. The error bars are
 estimated using the variance of the 
binomial law for the random variable $n_i$ leading to the following approximation for $p_i$, namely
 $n_i/N \pm (p_i(1-p_i)/N)^{1/2}$.

\section{Information based estimation of the average dissipation}
\begin{figure}[t!]
\includegraphics[width=\columnwidth]{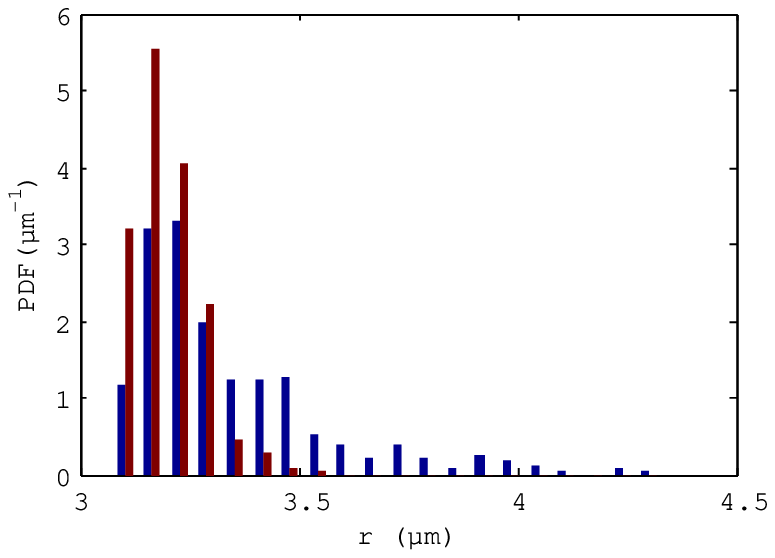}
\hspace{2mm}
\caption{\label{figS4}
Histogram of the probability distribution $p_F(t)$ (blue bars) and $p_R(\tau-t)$ (red bars) evaluated at the time $t=1.5$.}
\end{figure}

\begin{figure}[t!]
\includegraphics[width=\columnwidth]{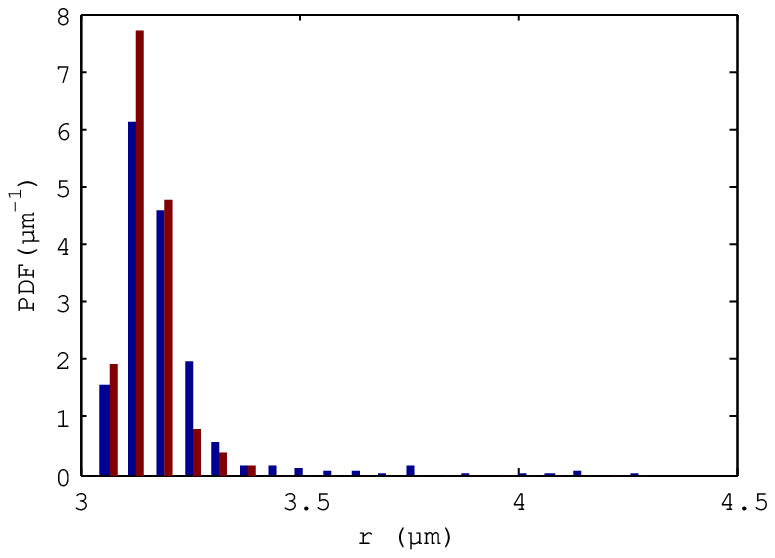}
\hspace{2mm}
\caption{\label{figS5}
Histogram of the probability distribution $p_F(t)$ (blue bars) and $p_R(\tau-t)$ (red bars) evaluated at the time $t=2$s.}
\end{figure}

The first method discussed in the main
text, to estimate the dissipation from the information content of the trajectories,
 relies on an estimate of the KL divergence of two probability distributions 
$p_F(t)$ and $p_R(\tau-t)$, evaluated at two points which are time-reversal symmetric in the protocol of
applied magnetic field. 
We recall the inequality which is used:
\beq
\beta \langle W_{diss}(\tau) \rangle \geq D(p_F(t) || p_R(\tau-t)).
\label{new Kawai form}
\eeq
In order to test this idea, we have evaluated the left and the right hand side
of this equation at different time $t$, which corresponds to the time of a specific point within the protocol 
of duration $\tau$ as defined in figure 1. Due to the symmetry of the protocol, the 
average dissipated work $\langle W_{diss}(\tau) \rangle$ is equal to the average work 
$\langle W(\tau) \rangle$ measured in the
first part of the paper.
For completeness, we include here two snapshots of the probabilities used to determine the KL estimate, 
at two specific times, namely at $t=1.5$s and $t=2$s.  On one hand, at the time $t=1.5$s, the KL divergence is close to its 
maximum as shown in Fig.~4 of the main text, and 
we can see in figure \ref{figS4} that indeed the two distributions differ significantly from each other at this time. 
On the other hand, at the time 
$t=2$s, the two distributions are closer to each other as seen in figure \ref{figS5} and as a result the KL estimate is lower.
It is interesting to note that in Fig.~4, the highest value of the KL divergence occurs in the second half of the protocol, 
only after the protocol has changed sign. This shows that the irreversibility of the evolution of the system is easier
 to determine from the information of the trajectories only after the system has reacted to a variation of the protocol  in time.

%

The second method to estimate dissipation from the information of the trajectories discussed in the main
text relies instead on an estimate of the KL divergence between an equilibrium $p_{eq}(t)$ 
and a non-equilibrium probability
distribution $p_{neq}(t)$. In such a formulation,  
 both distributions need to be evaluated at the current value of the control parameter at time $t$ which is the 
final time for the evaluation of the dissipated work $W_{diss}(t)$: 
\beq
\beta \langle W_{diss}(t) \rangle \geq D(p_{neq}(t) || p_{eq}(t)).
\label{KL_bound}
\eeq
In order to test this relation, we have again varied the time $t$ as before.
In such a case, since the protocol is no longer symmetric, i.e. it does not take the same value 
at the initial and final time used in the evaluation of $\langle W_{diss}(t) \rangle$,
one needs to take into account the contribution from the free energy. Thus, we evaluate
$\langle W_{diss}(t) \rangle$ from $\langle W(t) \rangle - \Delta F(t)$, where $\Delta F(t)=F(B(t))-F(B(0))$ 
is the equilibrium free energy difference evaluated with a protocol taken at time $t$ and at the initial time $0$.
In order to evaluate this free energy, we have calculated it numerically from the partition function as the
following 2D integral
 $F(B)=-k_B T \log Z(B)$, where 
 \beq
 Z(B) = \int d {\bf r} \exp \left( -\beta U(r,\theta,B) \right),
 \eeq
where the integrand contains the 2D interaction potential introduced in the first section of these notes.
At low field, the potential is not sufficiently confining and the integral over the distance between the beads $r$ 
needs to be regularized. In order to do this, we have introduced a cutoff which corresponds to
the maximum distance observed in the experiment between the beads, namely a few microns.
To summarize, the evaluation of $\langle W_{diss}(t) \rangle$ requires the experimentally determined 
values of the work on the 460 cycles used before, together with the above free energy difference evaluated
with the corresponding value of the protocol of magnetic field at the time $t$.
As expected from Eq. \ref{KL_bound}, the curve corresponding to  $\langle W_{diss}(t) \rangle$
lies above the one corresponding to the KL bound at all times except in a small region at very early time.
In this region, both $\langle W_{diss}(t) \rangle$ and its KL estimate are small and their precise evaluation is 
more difficult than at later times. 
We attribute the discrepancy seen at very early time to the fact that the equilibration at the initial time may 
not be perfect and 
the experimental probability distribution may differ slightly from $p_{eq}(\tau)$ which is evaluated using the
theoretical model for the interaction potential. 

\bibliographystyle{myapsrev}
\bibliography{anupam}